\documentclass[twocolumn,showpacs,preprintnumbers,amsmath,amssymb]{revtex4}
\usepackage{graphicx}
\usepackage{bm}%{amsmath}
\usepackage{dcolumn}
\makeatletter

\newcommand{\Rmnum}[1]{\expandafter\@slowromancap\romannumeral #1@}
\makeatother

\begin{document}
\title{ Low-momentum interactions with Brown-Rho-Ericson scalings\\
and the density dependence of the nuclear symmetry energy}
\author{Huan Dong and T.\ T.\ S.\ Kuo}
\affiliation{Department of Physics and Astronomy,
Stony Brook University, New York 11794-3800, USA}
\author{R. Machleidt}
\affiliation{Department of Physics,
University of Idaho, Moscow, Idaho 83844, USA}
\begin{abstract}
We have calculated the nuclear
symmetry energy $E_{sym}(\rho)$ up to densities of $4 \sim 5 \rho_0$
with the effects from the Brown-Rho (BR) and Ericson scalings for 
the in-medium mesons
included. Using the $V_{low-k}$ low-momentum interaction with and without
such scalings,  the equations of state (EOS) of symmetric and asymmetric 
nuclear matter have been calculated using a ring-diagarm formalism where
 the particle-particle-hole-hole ring diagrams are included to all orders.
The EOS for symmetric nuclear matter and neutron matter obtained with 
linear BR scaling are both overly stiff
compared with the empirical constraints of Danielewicz {\it et al.}
\cite{daniel02}.
In contrast, satisfactory results are obtained by either using the nonlinear
Ericson scaling or by adding a Skyrme-type 
three-nucleon force (TNF)
to the unscaled $V_{low-k}$ interaction.
 Our results for $E_{sym}(\rho)$ obtained 
with the nonlinear Ericson scaling are in good agreement 
with the empirical values of Tsang {\it et al.} \cite{tsang09} 
and Li {\it et al.} \cite{li05}, while those with TNF are slightly
below these values. For densities below the nuclear saturation density
$\rho_0$, the results of the above calculations are nearly equivalent
to each other and all in satisfactory agreement with the empirical values.

\end{abstract}
 \maketitle
\section{INTRODUCTION}

The nuclear matter symmetry energy  is an important  as well as 
very interesting
subject in  nuclear and astro-nuclear physics. As reviewed extensively 
in the literature 
\cite{baran05,steiner05,latti07,li08,ditoro08,sammarruca10,tsang09,xiao09},
it plays a crucial role in determining many important
nuclear properties, such as
the neutron skin of nuclear systems,  structure
of nuclei near the drip line, and neutron stars'  masses and radii.
It is especially of importance that constraints on 
the nuclear matter equation of state (EOS) \cite{daniel02} 
and the density ($\rho$) dependence
of the symmetry energy $E_{sym}(\rho)$ \cite{li05,tsang09}
up to $\rho \simeq 4\rho_0$  have been
experimentally extracted from heavy-ion collisions, 
$\rho_0$ being the saturation density
of symmetric nuclear matter. 
There have been a large number of theoretical derivations of $E_{sym}(\rho)$
using, for example, the Brueckner Hartree-Fock (BHF) 
\cite{bombaci91,zhi02,zhli06}, Dirac BHF \cite{sammarruca10,alonso03,krastev06,dalen07},
variational  \cite{wiringa88}, relativistic mean field (RMF)
\cite{lwchen07} and Skyrme HF \cite{lwchen05} many-body methods.
The results of these theoretical investigations 
have exhibited, however,
large variations for $E_{sym}(\rho)$.
Depending on the interactions and many-body methods used, they can give
either a `hard' $E_{sym}(\rho)$, 
in the sense that it increases monotonically
with $\rho$ up to $\sim 5 \rho_0$, or a `soft' one where
  $E_{sym}(\rho)$ arises to  a maximum value at $\rho \simeq 1.5 \rho_0$ and 
then descends
to zero at $\sim 3 \rho_0$ \cite{li08,xiao09}. It appears that 
the predicted behavior 
of $E_{sym}(\rho)$ may  depend importantly on the nucleon-nucleon (NN)
interactions and the many-body methods employed.

In the present work, we shall calculate the nuclear symmetry energy
using the low-momentum
interaction $V_{low-k}$ derived from realistic NN interactions $V_{NN}$
 using a renormalization group approach 
\cite{bogner01,bogner02,bogner03,bogner03b,jdholt}. To our knowledge, this 
renormalized interaction has not yet been applied to the study of $E_{sym}$.
As it is well known, most realistic $V_{NN}$  contain  hard cores, 
or strong short-range
repulsions. This feature makes these interactions not suitable
for being directly used in nuclear many-body calculations; they 
need to be `tamed' beforehand. For many years,
this taming is enacted by way of the BHF theory where $V_{NN}$ is converted
into the Brueckner $G$-matrix. 
A complication of the $G$-matrix is its  energy dependence (see e.g.
\cite{kuoma86}), making it
rather inconvenient for calculations. In the $V_{low-k}$ approach,
a different `taming' procedure is employed; it is performed by 
`integrating out' the high-momentum components of $V_{NN}$ beyond a 
decimation scale $\Lambda$. In this way, the resulting $V_{low-k}$ is
energy independent. Furthermore $V_{low-k}$ is nearly unique, 
namely the $V_{low-k}$s
deduced from various realistic $V_{NN}$ potentials
(such as \cite{cdbonn,nijmegen,argonne,mach89}) are nearly identical to each
other for decimation scale $\Lambda \simeq 2 fm^{-1}$ 
\cite{bogner03,bogner03b}.

Using this $V_{low-k}$ interaction, we shall first calculate the
equations of state (EOS)  $E(\rho,\alpha)$ for asymmetric nuclear  matter, 
from which $E_{sym}(\rho)$ can be obtained. 
Here $E$ is the ground-state energy per nucleon and  
$\rho$ is the total baryon density.
$\alpha$ is the isospin asymmetry parameter defined as
$\alpha=(\rho_n-\rho_p)/\rho$, where $\rho_n$ and $\rho_p$ denote, repspectively,
the neutron and proton density and $\rho=\rho_n+\rho_p$. 
Our EOS will be calculated  using a ring-diagram 
many-body method \cite{siu09,dong09,song87}.
As we shall discuss later, this method includes the particle-particle
hole-hole ($pphh$) ring diagrams to all orders. In comparison, only the
 diagrams with two hole lines are included in the familiar HF, BHF
and DBHF calculations. In other words, in these HF methods a closed
Fermi sea is employed while in the ring-diagram framework the effects
from the  fluctuations of the Fermi sea are taken into account by
including the $pphh$ ring diagrams to all orders. 
 
The nuclear symmetry energy $E_{sym}(\rho)$ is related to the asymmetric
nuclear matter EOS by
\begin{equation}
E(\rho,\alpha)=E(\rho,\alpha=0)+E_{sym}(\rho)\alpha^2
+O(\alpha^4).
\end{equation}
The contributions from terms
of  orders higher  than $\alpha^2$ are  usually negligibly small,
as illustrated by our results in section III. 
  With such contributions neglected, 
we have 
\begin{equation}
        E_{sym}(\rho)=E(\rho,1)-E(\rho,0).
\end{equation}
Then the symmetry energy is just given by the energy difference between
neutron
and symmetric nuclear matter. In calculating $E_{sym}(\rho)$,
the above EOS clearly play an important role. In our calculation, we shall
 require that the NN interaction and many-body
methods employed should give satisfactory results 
for $E(\rho,1)$ and $E(\rho,0)$ of, respectively, neutron and symmetric nuclear
matter.
The use of $V_{low-k}$ alone, however, has not been able to  
reproduce the empirical nuclear saturation properties, the 
predicted saturation density
and binding energy per particle being both too large compared with the
empirical values of 
$\rho_0\simeq 0.16 {\rm fm}^{-3}$ and  
$E \simeq -16  {\rm MeV}$ for symmetric nuclear matter \cite{siu09,dong09}. 
To improve the situation, it may be necessary to include the effects from
Brown-Rho (BR) scaling \cite{brownrho1,hatsuda,rapp} for the in-medium
mesons, or 
a three-nucleon force (TNF) \cite{bogner05}. BR scaling is suitable
only for the low density region; it suggests that the masses
of light vector mesons in medium are reduced `linearly'  with the density.
We consider here the EOS up to about $\sim 5 \rho_0$ and at such high
density the linear BR scaling is clearly not applicable. In the present 
work we shall adopt the nonlinear Ericson scaling \cite{ericson}
for the in-medium mesons 
and apply it to our $E_{sym}(\rho)$ calculations.
The effects from the linear BR and nonlinear Ericson scalings 
on the nuclear EOS and 
symmetry energy will be studied.

The organization of this paper is as follows. In section II we shall
briefly describe our derivation of the low-momentum interaction $V_{low-k}$
using a $T$-matrix equivalence approach. Some details about the calculation
of the EOS for asymmetric nuclear matter from this interaction with 
the $pphh$ ring diagrams summed
to all orders will also be presented. The Ericson scaling is a nonlinear
extension of linear BR scaling. The difference between them
will be addressed in this section. Our results will be presented and 
discussed in section III. A summary and conclsion is contained in 
section IV.

\section{FORMALISM}
 
We shall calculate $E_{sym}(\rho)$ using a low-momentum  ring-diagram approach
\cite{siu09,dong09,song87}, where the $pphh$ ring diagrams 
are summed to all orders
within a  model space of decimation scale $\Lambda$. 
In this approach, we employ the low-momentum interaction $V_{low-k}$ 
\cite{bogner01,bogner02,bogner03,bogner03b,jdholt}. Briefly speaking, this
interaction is obtained by solving the following $T$-matrix equivalence 
equations: 
\begin{multline}
T(k',k,k^2)=V_{\rm NN}(k',k) \\
+\frac{2}{\pi}\mathcal{P}\int_0^\infty
\frac{V_{\rm NN}(k',q)T(q,k,k^2)}{k^2-q^2}q^2dq,
\end{multline}
\begin{multline}
T_{\rm low-k}(k',k,k^2)=V_{\rm low-k}(k',k)\\
+\frac{2}{\pi}\mathcal{P}\int_0^\Lambda
\frac{V_{\rm low-k}(k',q)T_{\rm low-k}(q,k,k^2)}{k^2-q^2}q^2dq,
\end{multline}
\begin{equation}
 T(k',k,k^2)=T_{low-k}(k',k,k^2); (k',k) \leq \Lambda.
\end{equation}
In the above $V_{NN}$ represents a realistic NN interaction such as the
CDBonn potential \cite{cdbonn}.  $\mathcal{P}$ denotes principal-value 
integration 
and the intermediate state momentum \emph{q} is integrated 
from 0 to $\infty$ for  the whole-space $T$ and from 0 to $\Lambda$
for $T_{\rm low-k}$. The above $V_{low-k}$ preserves the low-energy
phase shifts (up to energy $\Lambda^2$) and the deuteron binding energy
of $V_{NN}$.  Since $V_{low-k}$ is obtained by integrating out the high-momentum
components of $V_{NN}$, it is a smooth `tamed' potential which is suitable
for being used directly in many-body calculations.

\begin{figure}[here]    
\scalebox{0.4}{\includegraphics{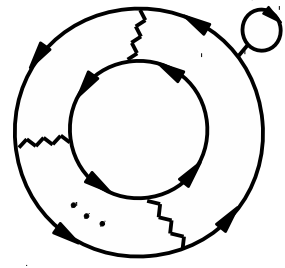}}
\caption{Sample ring diagram included in the equation of state $E(\rho,\alpha)$.
Each wave line represents a $V_{low-k}$ vertex. The HF one-bubble
insertions to the Fermion lines are included to all orders.}
\end{figure}
We use a ring-diagram method \cite{siu09,dong09,song87} to calculate the
nuclear matter {\rm EOS}. 
In this method, the ground-state energy is expressed as 
$E(\rho,\alpha)=E^{free}(\rho,\alpha)+ \Delta E(\rho,\alpha)$
where $E^{free}$ denotes the free (non-interacting) EOS
and $\Delta E$ is the energy shift due to the NN interaction.
In our ring-diagram approach, $\Delta E$ is 
is given by the all-order sum of the \emph{pphh} ring diagrams as illustrated
in Fig. 1. Note that we include three types of ring diagrams, the
proton-proton, neutron-neutron and proton-neutron ones. The proton and neutron
Fermi momenta are, respectively, 
$k_{Fp}=(3\pi^2\rho_p)^{1/3}$  
and $k_{Fn}=(3\pi^2\rho_n)^{1/3}$.  
With such ring diagrams summed to all orders,  we have
\cite{dong09,song87} 
\begin{multline}\label{eng}
\Delta E(\rho,\alpha)=\int_0^1 d\lambda
\sum_m \sum_{ijkl<\Lambda}Y_m(ij,\lambda) \\ \times Y_m^*(kl,\lambda) \langle
ij|V_{\rm low-k}|kl \rangle,
\end{multline}
where the transition amplitudes $Y$ are obtaind from a $pphh$ RPA equation
\cite{siu09,dong09,song87}.
Note that $\lambda$ is a strength parameter, 
integrated from 0 to 1. The above ring-diagram method reduces to the
usual HF method if only the first-order ring diagram 
is included. In this case, the above energy shift becomes 
$\Delta E(\rho,\alpha)_{HF}=\frac{1}{2}
\sum n_i n_j\langle ij|V_{\rm low-k}|ij \rangle$ where
$n_k$=(1,0) if $k(\leq,>)k_{Fp}$ for  proton 
and $n_k$=(1,0) if $k(\leq,>)k_{Fn}$ for  neutron.

It is well known that the use of the free-space $V_{NN}$ alone
 is not adequate for describing nuclear properties
at high densities. 
To satisfactorily describe such properties, one may need
to include the three-nucleon force \cite{bogner05} or the in-medium
modifications to the nuclear interaction.
In the present work, we shall employ in our EOS calculations
nuclear interactions which contain the in-medium modifications suggested
by the
 Brown-Rho (BR) \cite{brownrho1,hatsuda} and Ericson \cite{ericson} scalings.
These scalings are based on the relation \cite{nambu,brownrho1,hatsuda} 
 that hadron masses scale with the quark condensate 
$\langle \bar{q}q \rangle$ in medium as
\begin{equation}
     \frac{m^*}{m}=\left(\frac{\langle \bar{q}q(\rho)\rangle}
     {\langle \bar{q}q(0)\rangle}\right)^{1/3}
\end{equation}
where $m^*$ is the hadron mass in a medium of density $\rho$, and
\emph{m} is that in free space.  
The  quark condenstate $\langle \bar{q}q \rangle$ 
measures the chiral symmetry breaking, and 
its density dependence in the low-density limit is related
\cite{cohen,lutz} to the free $\pi N$ sigma term $\Sigma _{\pi N}$ by  
\begin{equation}
\frac{\langle \bar{q}q(\rho)\rangle}{\langle \bar{q}{q}(0)\rangle}
=1-\frac{\rho\Sigma_{\pi N}}{f_\pi^2m_\pi^2}
\end{equation}
where  $f_\pi=93 {\rm MeV}$ is the pion
decay constant and $\Sigma_{\pi N}=45 \pm 7 {\rm MeV}$ \cite{gasser}. 
Applying the above scaling to   mesons in  low-density nuclear medium, 
one has the 
linear scaling \cite{hatsuda}
\begin{equation}
     \frac{m^*}{m}=1-C\frac{\rho}{\rho_0}
\end{equation}
where $m^*$ and $m$ are, respectively, the in-medium and free meson mass,
and \emph{C} is a constant of value $\sim 0.15$. The above scaling
will be  referred to as the linear BR scaling.
Nucleon-nucleon interactions are mediated by meson exchanges,
and clearly  the in-medium modifications of meson masses can significantly
alter the NN interaction. These modifications
could arise from the partial restoration of chiral symmetry at finite
density/temperature or from traditional many-body effects.
Particularly important
are the vector mesons, for which there is now evidence from both theory
\cite{hatsuda, harada, klingl} and experiment
\cite{metag, naruki}
that the masses may decrease by approximately $10-15\%$ at normal nuclear
matter density and zero temperature. 
(Pions are not scaled because they are protected by chiral symmetry.)
Density-dependent nuclear interactions obtained by applying the above scaling 
 to the light mesons ($\omega$, $\rho$
and $\sigma$)  which mediate the NN potential have been employed in studying
the properties of nuclear matter
\cite{rapp,holt07,siu09,dong09} and the $^{14}C\rightarrow ^{14}N$
$\beta$-decay \cite{jwholt08}.

We are interested in the EOS and $E_{sym}$ up to  densities
 as high as $\rho \simeq 5 \rho_0$, and at such high densities 
 the above linear scaling is clearly not suitable.
How to scale the mesons in such high density region is 
still by and large  uncertain.
We shall adopt here the  Ericson scaling \cite{ericson} which is an
extension of the BR scaling.
In this scaling,  a 
new relation for the quark condensate $\langle \bar{q}q \rangle$
based on chiral symmetry breaking
 is employed, namely
\begin{equation}
\frac{\langle\bar{q}q(\rho)\rangle}{\langle\bar{q}q(0)\rangle}
=\frac{1}{1+\frac{\rho \Sigma_{\pi N}}{f_\pi^2m_\pi^2}}.
\end{equation}
Note that this relation agrees with the linear  scaling relation of Eq.(8)
for small $\rho$.
The above scaling suggests a  non-linear  scaling for meson mass
\begin{equation}
\frac{m^*}{m}=\left(\frac{1}{1+D\frac{\rho}{\rho_0}}\right)^{1/3}
\end{equation}
with $D=\frac{\rho_0 \Sigma_{\pi N}}{f_\pi^2m_\pi^2}$, 
and we shall  refer to this scaling as the nonlinear Brown-Rho-Ericson (BRE)
 scaling.
Using the empirical values for 
($\Sigma _{\pi N},~\rho_0,~f_{\pi},~m_{\pi}$), we have 
\emph{D}=  0.35$\pm$0.06.
In the present work, we shall employ the one-boson exchange
BonnA potential \cite{mach89}
with its ($\rho,~\omega,~\sigma$)
mesons scaled using both the linear (Eq.(9)) 
and nonlinear (Eq.(11)) scalings. This potential is chosen because
it has a relatively simple structure which is convenient for scaling
its meson parameters.

\section{Results and discussions}

Using both the unscaled and scaled BonnA potentials, we  first calculate 
the ring-diagram EOS for symmetric nuclear matter
to investigate if they can give saturation properties in good agreement
with the empirical values. We employ the low-momentum interactions
$V_{low-k}$ from these potentials using a decimation $\Lambda=3.0 fm ^{-1}$, 
which is chosen because  we are to study the EOS up to high densities
of $\sim 5 \rho_0$. As shown in Fig. 2,  the EOS 
(labelled '$V_{low-k}$ alone') calculated with 
the unscaled potential saturates at $k_F\simeq 1.8 fm^{-1}$, which is
too large compared with the empirical value, and it also overbinds
 nuclear matter. We then repeat the calculation including
the medium modifications from the BR scalings. For the linear BR scaling
(Eq.(9)),
we have used $C_{\omega}$=0.128, $C_{\rho}$=0.113 and $C_{\sigma}$=0.102.
These parameters are chosen so as to have satisfactory saturation
properties, namely they give $E_0/A \simeq$-15.5 MeV and $\rho _0 \simeq$
0.17 $fm^{-3}$. In Fig. 2 we also present our results obtained with the
nonlinear BRE scaling (Eq.(11)) using parameters $D_{\omega}=D_{\rho}$=0.40
and $D_{\sigma}$=0.30. They were chosen to provide satisfactory results
for $E_0/A$ and $\rho _0$. It is of interest that 
for densities $\lesssim \rho_0$
the EOS given by the linear BR and nonlinear BRE scalings are practically
eqivalent to each other. From Eqs.(7-11), we see that the
parameters
$D$ and $C$ obtained from the density dependence of quark condensates
is  $0.29 \lesssim D \lesssim 0.41$ and $C \simeq D/3$. It is noteworthy
that the  $C$ and $D$ parameters we have employed in the EOS calculations
agree well with the above theoretical values. 

As also seen from Fig. 2, the above equivalence begins to disappear
for densities larger than $\rho_0$. There the EOS given by the linear
scaling is much stiffer than that given by the nonlinear one; the difference
between them becomes larger and larger as density increases.
In addition to the above two EOS, we have also calculated an EOS using
the interaction given by the sum of the unscaled $V_{low-k}$ and an
empiriral Skyrme three-nucleon force (TNF).
The well-known emipirical Skyrme force \cite{skyrme} 
is of the form
\begin{equation}
V_{\rm Skyrme}=\sum_{i<j} V(i,j) + \sum_{i<j<k} V(i,j,k),
\end{equation}
where \emph{V}(\emph{i,j}) is a two-nucleon momentum dependent interaction,
and V(i,j,k) is  a zero-range three-nucleon interaction which has played
an indispensible role for nuclear saturation.
 For nucleons in a nuclear medium of
density $\rho$, this three-nucleon force becomes a 
density-dependent two-nucleon
force commonly  written as
\begin{equation}
V_{\rho}(i,j)=\frac{t_3}{6}\rho \delta(\vec r_i-\vec r_j).
\end{equation}
 In Fig. 2 the EOS labelled  '$V_{low-k}$ with TNF' is obtained
using the combined interaction of $V_{low-k}$ (unscaled) and  $V_{\rho}$. 
The parameter
$t_3$ is adjusted so that the resulting EOS gives satisfactory
saturation properties for symmetric nuclear matter. The EOS
shown has
$t_3$=2000 MeV-$fm^{6}$.

\begin{figure}[tbh]
\scalebox{0.42}{
\includegraphics[angle=-90]{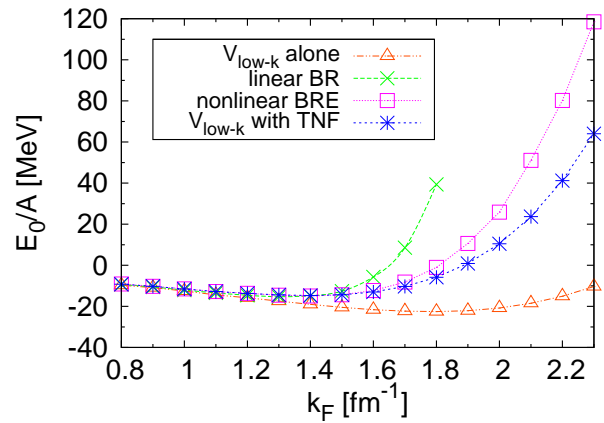} 
}
\caption{Ring-diagram {\rm EOS}s calculated with the $V_{low-k}$ 
interaction alone,  with the linear BR scaling of Eq.(9), 
with the nonlinear BRE scaling of Eq.(11), and  with the addition of
a Skyrme-type
three-nucleon force (TNF).}
\end{figure}
 
It is of interest that the above three EOS (linear BR and nonlinear BRE, TNF)
are nearly identical for
densities $\lesssim \rho_0$, but they deviate from each other
with increasing densities.
Without  experimental guidelines about the nuclear matter EOS
above $\rho_0$, it would be difficult to determine
which of these three EOS has the correct high density behavior.
Fortunately, heavy-ion collision experiments conducted during the last 
several years have provided us with  constraints of
the EOS at high densities.
 Danielewicz \emph{et al.} \cite{daniel02} have obtained
 a constraint on the EOS for symmetric nuclear matter
 of densities between $2\rho_0$ and $4.5\rho_0$, 
as shown by the red solid-line box in Fig. 3. 
Comparing our three {\rm EOS}s with their
constraint,  the linear {\rm BR EOS} is clearly not consistent with the
constraint and should be ruled out. This linear scaling is suitable for low
densities, but definitely needs modification at high densities. It is primarily
for this purpose that we have considered the nonlinear scaling. 
 As displayed in Fig. 3, the EOS with the nonlinear BRE scaling is 
in much better agreement
 with the constraint than the linear BR one. It satisfies the constraint
well except being slightly above the constraint at densiies near 
$\sim 4.5 \rho_0$.
It is of interest that the EOS using $V_{low-k}$ with the Skyrme-type 
TNF exhibits even better agreement
with the constraint. 

\begin{figure}[here]
\scalebox{0.42}{
\includegraphics[angle=-90]{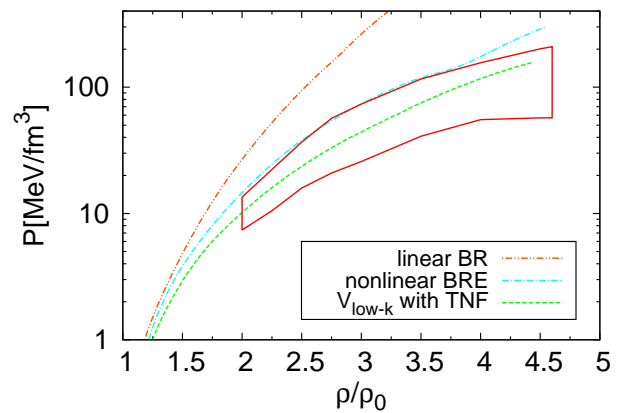} 
}
\caption{Comparison of the calculated equations of state for symmetric nuclear
matter with the constraint (solid-line box) of Danielewicz \cite{daniel02}.
See text for more explanations.}
\end{figure}

So far we have studied the effects of the BR scalings and the TNF three-nucleon
force on the EOS for symmetric nuclear matter.
The neutron matter {\rm EOS} is also 
an interesting and important topic \cite{pandar,babrown}.
It plays a crucial role in determining the nuclear symmetry energies
as well as the properties of neutron stars.
It should be of interest to study also the effects
of the above BR/BRE scalings and the TNF force on the EOS of neutron matter.
Using the same $V_{low-k}$ ring-diagram framework employed for
symmetric nuclear matter and  the same $C$, $D$ and $t_3$ parameters,
 we have  
caculated the neutron matter {\rm EOS} up to 4.5$\rho_0$.
Our calculated neutron-matter EOS are displayed in Fig. 4.
Danielewicz \emph{et al.} \cite{daniel02} have given two 
different constraints for the neutron matter {\rm EOS}: a stiff
one (upper black solid-line box) 
and a  soft one (lower red solid-line box) 
 which are both displayed in Fig. 4.
As we can see, the linear {\rm BR EOS} is 
again producing too much pressure.
The nonlinear {\rm BRE EOS} agrees well with  the 
stiff constraint (upper box) while the {\rm TNF EOS} is 
fully within the soft constraint box. 
To further test these
two EOS (nonlinear BRE and TNF), it would be very helpful
to have narrower experimental
constraints on the neutron matter EOS.

\begin{figure}[htb]
\scalebox{0.42}{
\includegraphics[angle=-90]{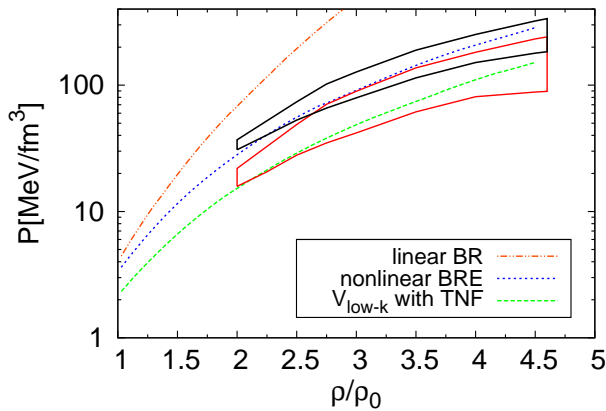} 
}
\caption{Comparison of the calculated equations of state for neutron-matter 
 with the
constraints of Danielewicz \cite{daniel02}. See text for more explanations.}
\end{figure}

The symmetry energy $E_{sym}$ is a  topic of much current interest,
and extensive studies have been carried out to extract
its  density dependence  from heavy-ion collision 
experiments \cite{tsang09,li05}.
Based on such experiments, Li \emph{et al.} \cite{li05} suggested an 
empirical relation
\begin{equation} 
E_{sym}(\rho) \approx 31.6(\rho/\rho_0)^\gamma;~ \gamma=0.69-1.1,
\end{equation}
for constraining the density dependence of the symmetry energy. 
 Also based on such experiments, Tsang \emph{et al.} \cite{tsang09} 
recently proposed a new empirical relation for the symmetry energy, namely
\begin{equation}
E_{sym}(\rho)=\frac{C_{s,k}}{2}\left(\frac{\rho}{\rho_0}\right)^{2/3}+
\frac{C_{s,p}}{2}\left(\frac{\rho}{\rho_0}\right)^{\gamma_i}
\end{equation} 
where $C_{s,k}=25 {\rm MeV}$, $C_{s,p}=35.2 {\rm MeV}$ 
and  $\gamma_i \approx 0.7$. It should be useful and of interest
to check if our calculated $E_{sym}(\rho)$ is consistent with 
the  above relations.

\begin{figure}[here]
\scalebox{0.40}{
\includegraphics[angle=-90]{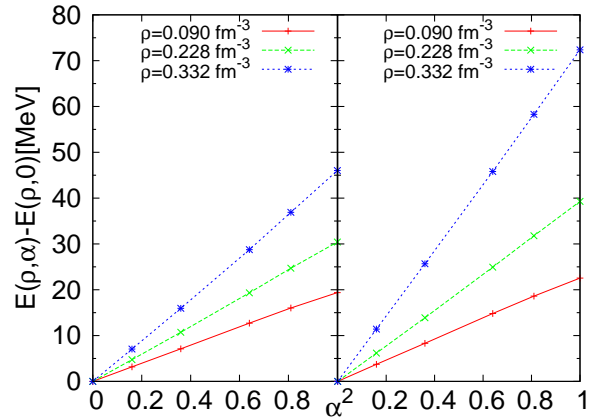} 
}
\caption{Ring-diagarm equations of state of asymmetric nuclear matter.
See text for more explanations. }
\end{figure}

Using the ring-diagram framework described earlier, we have calculated
the ground-state energy $E(\rho,\alpha)$ for asymmetric nuclear matter.
(Recall that the asymmetry parameter is $\alpha = (\rho_n-\rho_p)/\rho$.)
Some representative results are shown in Fig. 5:  the results in the left
panel are obtained with the '$V_{low-k}$ with TNF' interaction while for the right
panel the 'nonlinear BRE' interaction is used. As seen, $E(\rho,\alpha)$ varies
with $\alpha ^2$ almost perfectly linearly, for a wide range of $\rho$.
(Note that in Fig. 5 we plot the energy difference 
$E_{sym}(\rho,\alpha)-E_{sym}(\rho,0)$.)
This is a desirable and remarkable result, indicating that 
our ring-diagram symmetry energy 
can be accurately obtained  from the simple relation given by Eq.(2), 
namely the energy
 difference between neutron and symmetric nuclear matter.

In Fig. 6, the 'shaded area' represents  the empirical constraint, Eq.(14), 
of Li {\it et al.} \cite{li05}. As seen, there are large uncertainties
in the high-density region. The empirical relation Eq.(15) of
 Tsang {\it et al.} \cite{tsang09} is given by the 'second curve
from bottom' in the figure. As seen, the density dependence of this
relation is slightly below the softest limit (lower boundary of the
shaded area) of Eq.(14). Our 'nonlinear BRE' results are 
in the middle of the shaded area, in good agreement with the empirical
constraint of \cite{li05}. Our results with the TNF force are 
below the empirical ones of both \cite{li05} and  \cite{tsang09},
giving a softer density dependence than both.
It may be  noticed that
for densities ($\rho \lesssim \rho_0$), the calculated and empirical results
are all in good agreement with each other.
The symmetry energies given by them  at $\rho_0$ are all close to 
$\sim 30 {\rm MeV}$, which is also the only well determined empirical value.
Furthermore, our calculated symmetry energies all increase 
monotonically with density.
We have required our nuclear matter EOS to satisfy certain empirical
 constraints, and with such requirements it may be difficult
 for our present calculations to have a soft 
 $E_{sym}(\rho)$ as soft as the supersoft one of \cite{xiao09}
which saturates at density near $\sim 1.5 \rho_0$.

We have found that our symmetry energies can be well fitted by expressions
of the same forms as Eqs.(14) and (15), with the exponents 
$\gamma$ and $\gamma _i$ treated
as parameters. In Table I, we compare the exponents determined from
our results with the empirical ones of \cite{li05} and \cite{tsang09}.
The $\gamma$ exponent given by the nonlinear BRE scaling is in good agreement
 with the empirical values of \cite{li05}. 
The empirical $\gamma _i$ of \cite{tsang09} is, however, 
about half-way between the $\gamma _i$
obtained with 'nonlinear BRE' and that with 'TNF'.

\begin{figure}[here]
\scalebox{0.42}{
\includegraphics[angle=-90]{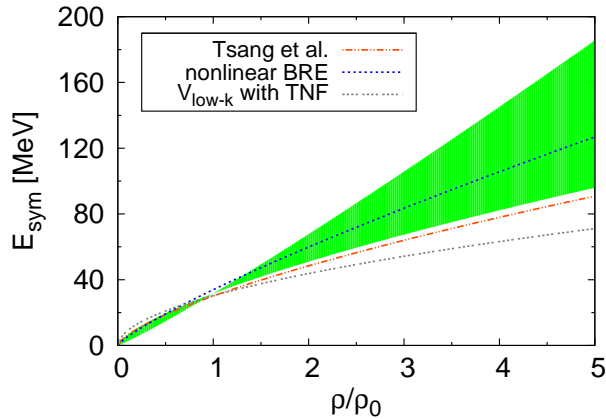} 
}
\caption{Comparison of the density dependence of our calculated
symmetry energies with the empirical results of \cite{tsang09}
(dot-dash line) and \cite{li05} (shaded area).}
\end{figure}
\begin{table}[here]
\caption{Comparison of the density exponents for the nuclear symmetry energy 
$E_{sym}(\rho)$. The exponents $\gamma$ and $\gamma _i$ are defined
respectively in Eqs. (14) and (15).}
%\centering
\begin{tabular}{|c|c|c|} \hline
        & $\gamma$ & $\gamma_i$\\  \hline
Li \emph{et al.} \cite{li05}    &  0.69-1.1 &  \\ \hline
Tsang \emph{et al.} \cite{tsang09}  &  & 0.7 \\ \hline
non-linear BRE        &  0.82 & 1.04  \\ \hline
TNF        &  0.53 & 0.43  \\ \hline
\end{tabular}
\label{table:fitted sym}
\end{table}

\section{SUMMARY AND CONCLUSION}
Employing the $V_{low-k}$ low-momentum interactions, we have calculated
the nuclear symmetry energy $E_{sym}(\rho)$ up to a density of
$\sim 5 \rho_0$ using a ring-diagram framework
 where \emph{pphh} ring diagrams are summed to all orders.
We first calculate the EOS for symmetric nuclear matter and neutron
matter and compare our results with the corresponding 
empirical constraints
of Danielewicz {\it et al.} \cite{daniel02}. To have satisfactory
agreements with such constraints, we have found it necessary
to include certain medium corrections to the free-space NN interations.
In other words, the effective NN interactions in medium are different
from those in free space, and when using them in nuclear many-body problems
it may be necessary to include the renormalization effects due to the
presence of other nucleons.
We have considered several methods to incorporate
such medium corrections. Although the nuclear matter saturation properties
can satisfactorily be reproduced by including the medium corrections
from the well-known linear Brown-Rho scaling
for the in-medium mesons, this scaling produces an EOS which is too stiff
compared with the Danielewicz constraints. We have found that the
EOS obtained with the nonlinear
Brown-Rho-Ericson  scaling 
are in good agreement with the Danielewicz constraints. 
We have considered another method to  render the effective interaction
density dependent, namely adding a Skyrme-type three-nucleon force (TNF)
to the unscaled $V_{low-k}$ interaction. The EOS so obtained are also
 in good agreement with the Danielewicz
constraints, but the resulting neutron matter EOS  is significantly
softer than that with the nonlinear scaling.  
The three methods (linear and nonlinear scalings, and TNF)  all have 
 reproduced well 
the empirical saturation  properties of nuclear matter ($\rho_0 \approx 0.17 
{\rm fm}^{-3}$ and $E_0/A \approx -15 {\rm MeV}$), but their results at
high densities are different. We have determined the scaling parameters
$C$ (linear BR scaling) and $D$ (nonlinear BRE scaling) 
by fitting the above saturation properties.
It is encouraging that the results, ($0.102\lesssim C \lesssim 0.128$)
and ($0.30\lesssim D \lesssim 0.40$), so obtained are actually 
in good agreement
with the theoretical result $D\simeq 0.35 \pm 0.06 \simeq 3C$ given
by Eqs.(7-11). 

 Including the above medium modifications, we proceed to calculate
the nuclear symmetry energies. We have found that the $E_{sym}(\rho,\alpha)$
given by our asymmetric ring-diagram calculations depends on
$\alpha ^2$ almost perfectly linearly. This is a rather surprising and useful
result, suggesting that the symmetry energy can be reliably
obtained from the simple energy difference between symmetric nuclear matter
and neutron matter. Our symmetry energies obtained with the
nonlinear BRE scaling agree well with the empirical constraints
of \cite{li05}, and are slightly above the empirical values of
\cite{tsang09}. Our results with the TNF force is slightly below
the empirical results of both \cite{li05} and \cite{tsang09}.
The non-linear Ericson scaling has given satisfactory results for
the equations of states of nuclear matter and nuclear symmetry
energies up to a density of $\sim 5 \rho_0$. We believe this scaling provides
a suitable extension of the linear BR scaling to moderately high densities
of $\lesssim 5 \rho_0$.
Our calculated $E_{sym}(\rho)$ all increase monotonically
with $\rho$ up to $\sim 5 \rho_0$. It may be of  interest to carry out
further studies about the possibility
of obtaining a supersoft symmetry energy which may saturate at some 
low density of $\sim 1.5\rho_0$ \cite{xiao09}.

{\bf Acknowledgement}
We thank Professor Danielwicz for sending us the experimental data, 
and G.E. Brown and E. Shuryak for many helpful discussions.
This work is supported in part by the U.S. 
Department of Energy under Grant Nos. DE-FG02-88ER40388 
and DE-FG02-03ER41270 (R.M.), and the U.S.
National Science Foundation under Grant No. PHY-0099444.

\end{document}